\documentclass[conference]{IEEEtran}
\usepackage{graphicx}
\usepackage{amsmath}
\usepackage{amssymb}
\usepackage{algorithmic}
\usepackage{algorithm}
\usepackage{tikz}
\usepackage{cite}
\usepackage{tikz}
\usetikzlibrary{arrows.meta, positioning, calc, fit, backgrounds}
\usepackage[nolist]{acronym}

\begin{document}
%
% work title
\title{Budget-Aware Federated Dual-Side Channel Estimation for Hybrid mmWave Massive MIMO}

\author{
    \IEEEauthorblockN{
        Jiawei~Chen\IEEEauthorrefmark{1}, 
        Ruining~Fan\IEEEauthorrefmark{1}, 
        Mouli~Chakraborty\IEEEauthorrefmark{2}, 
        Avishek~Nag\IEEEauthorrefmark{3}, 
        and~Anshu~Mukherjee\IEEEauthorrefmark{1}\IEEEauthorrefmark{4}
    }
    \IEEEauthorblockA{
        \IEEEauthorrefmark{1}School of Electrical and Electronic Engineering, University College Dublin, Ireland\\
        \IEEEauthorrefmark{2}School of Computer Science and Statistics, Trinity College Dublin, Ireland\\
        \IEEEauthorrefmark{3}School of Computer Science, University College Dublin, Ireland\\
        \IEEEauthorrefmark{4}Beijing-Dublin International College, Beijing University of Technology, Chaoyang, Beijing, China\\
        Email: \{jiawei.chen, ruining.fan\}@ucdconnect.ie, moulichakraborty@ieee.org, \\
        avishek.nag@ucd.ie, anshu.mukherjee@ieee.org
    }
}

\maketitle


\begin{abstract}
This work studies communication-constrained federated dual-side channel state information (CSI) estimation in hybrid millimeter-wave (mmWave) massive multiple input multiple output (MIMO) systems. Accurate CSI recovery is challenging because hybrid beamforming yields compressed and noisy observations, while repeated model exchange in federated learning (FL) makes communication efficiency strongly dependent on estimator size. Rather than developing a new federated optimization algorithm, we focus on estimator design under standard federated averaging (FedAvg) and study how to use a limited parameter budget effectively under repeated model exchange. Based on this perspective, we propose a budget-aware recalibrated refinement network (BARRNet), which combines a compact residual backbone with lightweight channel-wise recalibration for dual-side CSI refinement. Simulation results show that BARRNet achieves a better normalized mean squared error (NMSE)--communication tradeoff than the backbone-only control and heavier convolutional neural network (CNN) baselines. At 5 dB SNR, for the -13 dB DL NMSE target, it reduces the cumulative communication required by 28.9\% relative to the architecture-matched backbone-only control. These results indicate that communication-efficient federated CSI estimation depends not only on model compactness, but also on how limited model capacity is used under repeated model exchange.
\end{abstract}


\section{Introduction}

\ac{mmWave} massive \ac{MIMO} is a key technology for high-capacity wireless communications due to its large bandwidth and spatial multiplexing capability. Fully digital beamforming requires one \ac{RF} chain per antenna, resulting in high hardware cost and power consumption in large arrays. Hybrid analog/digital beamforming provides a practical alternative by using fewer RF chains while retaining most of the beamforming gain \cite{Low_complexity_Hybrid_Precoding_2014,Hybrid_Precoding_2014}. However, accurate \ac{CSI} remains essential, and channel estimation is challenging because observations are compressed by the hybrid \ac{RF} architecture and corrupted by noise.

Recent learning-based methods have shown strong potential for channel estimation, including dual-side \ac{CSI} refinement \cite{dong2019deep,soltani2019deep,zhang2017beyond}. However, most existing approaches are trained centrally, assuming that channel samples collected at user devices can be aggregated at a server. In practical wireless networks, channel data are naturally distributed across users, and direct sharing may incur excessive communication overhead or raise privacy concerns.

\ac{FL} addresses this issue by keeping local data at clients and exchanging only model parameters during training \cite{konevcny2016federated,elbir2021federated}. However, repeated model exchange still incurs substantial overhead, especially on the uplink. Under fixed-precision transmission, this overhead is largely determined by the number of trainable parameters. Existing communication-efficient \ac{FL} studies have mainly focused on optimization improvement and update reduction through quantization or compression \cite{alistarh2017qsgd,chen2021communication,konevcny2016federated}. By comparison, the allocation of a limited estimator parameter budget under repeated model exchange has been less explicitly studied in federated \ac{CSI} estimation. Unlike prior \ac{FL}-based channel estimation studies, including recent dual-side massive MIMO and RIS-aided settings \cite{han2025federated,qiu2024federated}, this work focuses on estimator-level parameter-budget allocation under repeated model exchange. In other words, most prior communication-efficient FL studies focus on reducing the cost of transmitting model updates, whereas this work focuses on how the estimator itself should be designed when every trainable parameter is repeatedly exchanged.

In this work, we study federated dual-side \ac{CSI} estimation from a budget-aware estimator design perspective. Rather than modifying the federated optimization algorithm, we fix standard \ac{FedAvg} and focus on estimator design under repeated parameter transmission. For the structured observations considered here, communication efficiency depends not only on keeping the estimator compact, but also on allocating limited parameters to the components that contribute most to performance.

Based on this perspective, we propose \ac{BARRNet}, which refines structured downlink and uplink observations using a compact residual backbone enhanced by a lightweight channel-wise recalibration module. An architecture-matched backbone-only control is used to isolate whether the gain comes from improved parameter budget allocation rather than merely adding parameters.

The main contributions of this work are summarized as follows:
\begin{itemize}
    \item We construct structured downlink and uplink observations from compressed hybrid measurements, enabling a unified learning-based \ac{CSI} refinement pipeline for dual-side channel estimation.
    \item We formulate communication-constrained federated dual-side \ac{CSI} estimation as an estimator-level parameter-budget-allocation problem under repeated model exchange and develop \ac{BARRNet} as a lightweight communication-aware refinement architecture under standard \ac{FedAvg}.
    \item We provide architecture-matched evidence that lightweight channel-wise recalibration uses a limited parameter budget more effectively than uniform backbone scaling and yields a better NMSE--communication tradeoff.
\end{itemize}

 

\begin{figure*}[t]
    \centering
    \includegraphics[width=0.8\textwidth]{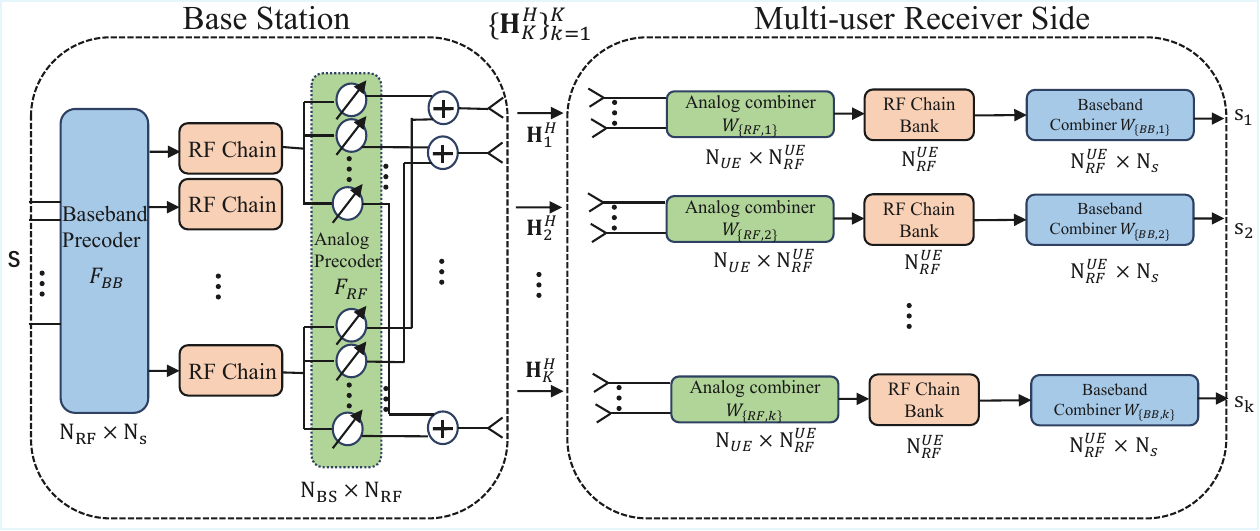}
    \caption{Hybrid multi-user massive MIMO system with dual-side hybrid beamforming. The BS performs digital baseband precoding followed by analog RF precoding, and transmits to multiple users through channels $\{\mathbf{H}^H_k\}_{k=1}^{K}$. Each user employs analog RF combining followed by baseband combining.}
    \label{fig:hybrid_system}
\end{figure*}

\section{System Model}

As illustrated in Fig.~\ref{fig:hybrid_system}, we consider a multi-user \ac{mmWave} massive \ac{MIMO} system with dual-side hybrid beamforming, where the \ac{BS} employs digital baseband precoding and analog \ac{RF} precoding, and each user employs analog RF combining followed by baseband combining. The \ac{BS} is equipped with $N_{\mathrm{BS}}$ antennas and $N_{\mathrm{RF}}$ \ac{RF} chains, and serves $K$ users. Each \ac{UE} is equipped with $N_{\mathrm{UE}}$ antennas and $N_{\mathrm{RF}}^{\mathrm{UE}}$ \ac{RF} chains. A fully connected hybrid structure is adopted, where the \ac{BS} employs the analog precoder $\mathbf{F}_{\mathrm{RF}}$ and the $k$th \ac{UE} employs the analog combiner $\mathbf{W}_{\mathrm{RF},k}$. Each \ac{RF} chain is connected to all antennas through phase shifters. The analog precoder and combiner satisfy the constant modulus constraints
\begin{equation}
|[\mathbf{F}_{\mathrm{RF}}]_{i,j}|=\frac{1}{\sqrt{N_{\mathrm{BS}}}}, \qquad
|[\mathbf{W}_{\mathrm{RF},k}]_{m,n}|=\frac{1}{\sqrt{N_{\mathrm{UE}}}}.
\end{equation}

We consider a narrowband block-fading mmWave channel. Let $\mathbf{H}_k\in\mathbb{C}^{N_\mathrm{BS}\times N_\mathrm{UE}}$ denote the uplink channel from the $k$-th \ac{UE} to the \ac{BS}. Using the geometric Saleh--Valenzuela model, it is given by\cite{Hybrid_Precoding_2014}
\begin{equation}\label{eq:channel_matrix_short}
\mathbf{H}_k
=
\sqrt{\frac{N_\mathrm{BS}N_\mathrm{UE}}{L}}
\sum_{l=1}^{L}
\alpha_{k,l}\,
\mathbf{a}_{BS}(\varphi_{k,l})
\mathbf{a}_{UE}^{H}(\vartheta_{k,l}),
\end{equation}
where $(\cdot)^{H}$ denotes the Hermitian transpose, $L$ is the number of propagation paths, $\alpha_{k,l}$ is the complex gain of the $l$th path, and $\varphi_{k,l}$ and $\vartheta_{k,l}$ denote the \ac{AoA} and \ac{AoD}, respectively. Here, $\mathbf{a}_{BS}(\varphi_{k,l})\in\mathbb{C}^{N_{\mathrm{BS}}\times 1}$ and $\mathbf{a}_{UE}(\vartheta_{k,l})\in\mathbb{C}^{N_{\mathrm{UE}}\times 1}$ denote the \ac{BS} and \ac{UE} array response vectors, respectively. Under \ac{TDD} operation, the corresponding downlink channel is $\mathbf{H}_k^H$. 

Since hybrid beamforming compresses the received measurements into a lower dimensional \ac{RF} domain, direct \ac{CSI} recovery is difficult. We therefore construct structured observations for both downlink (DL) and uplink (UL) transmission so that they become suitable for learning based refinement \cite{Low_complexity_Hybrid_Precoding_2014,Hybrid_Precoding_2014,han2025federated}.

For DL training, standard \ac{DFT} codebooks are used at both
the \ac{BS} and the \ac{UE} \cite{han2025federated}. Let
$\bar{\mathbf{F}}_{\mathrm{BS}}
\in\mathbb{C}^{N_{\mathrm{BS}}\times N_{\mathrm{BS}}}$
and
$\bar{\mathbf{W}}_{\mathrm{UE},k}
\in\mathbb{C}^{N_{\mathrm{UE}}\times N_{\mathrm{UE}}}$
denote the corresponding normalized \ac{DFT} codebook matrices
at the \ac{BS} and the $k$th \ac{UE}, respectively. The received
DL pilot observation is
\begin{equation}
\mathbf{Y}_k^{\mathrm{DL}}
=
\bar{\mathbf{W}}_{\mathrm{UE},k}^{H}
\mathbf{H}_k^{H}
\bar{\mathbf{F}}_{\mathrm{BS}}
+
\mathbf{Z}_k^{\mathrm{DL}},
\end{equation}
where
$\mathbf{Z}_k^{\mathrm{DL}}
\in\mathbb{C}^{N_{\mathrm{UE}}\times N_{\mathrm{BS}}}$
is additive Gaussian noise. Projecting the observation back
through the conjugate \ac{DFT} bases yields
\begin{equation}\label{eq:Rk_short}
\mathbf{R}_k
=
\bar{\mathbf{W}}_{\mathrm{UE},k}
\mathbf{Y}_k^{\mathrm{DL}}
\bar{\mathbf{F}}_{\mathrm{BS}}^{H}
=
\mathbf{H}_k^{H}
+
\widetilde{\mathbf{Z}}_k^{\mathrm{DL}},
\end{equation}
where
$\widetilde{\mathbf{Z}}_k^{\mathrm{DL}}
=
\bar{\mathbf{W}}_{\mathrm{UE},k}
\mathbf{Z}_k^{\mathrm{DL}}
\bar{\mathbf{F}}_{\mathrm{BS}}^{H}$
and
$\mathbf{R}_k\in
\mathbb{C}^{N_{\mathrm{UE}}\times N_{\mathrm{BS}}}$
is the structured noisy DL observation.

For UL training, each \ac{UE} transmits an orthogonal pilot
matrix
$\mathbf{P}_k\in\mathbb{C}^{N_{\mathrm{UE}}\times T}$,
where
$\mathbf{P}_k\mathbf{P}_k^{H}=\mathbf{I}$
and
$\mathbf{P}_k\mathbf{P}_j^{H}=\mathbf{0}$ for $k\neq j$.
To probe the \ac{BS} array with only $N_{\mathrm{RF}}$ RF chains,
the \ac{BS} employs $D$ successive analog combiners, giving
\[
\mathbf{F}_{\mathrm{eq}}
=
[\mathbf{F}_1,\ldots,\mathbf{F}_D]
\in
\mathbb{C}^{N_{\mathrm{BS}}\times DN_{\mathrm{RF}}}.
\]
The aggregated UL pilot observation is
\begin{equation}
\mathbf{Y}^{\mathrm{UL}}
=
\mathbf{F}_{\mathrm{eq}}^{H}
\sum_{k=1}^{K}\mathbf{H}_k\mathbf{P}_k
+
\mathbf{Z}^{\mathrm{UL}},
\end{equation}
where
$\mathbf{Z}^{\mathrm{UL}}
\in\mathbb{C}^{DN_{\mathrm{RF}}\times T}$
is additive Gaussian noise. Pilot correlation and spatial
back-projection give
\begin{equation}\label{eq:Gk_short}
\mathbf{G}_k
=
\mathbf{F}_{\mathrm{eq}}
\mathbf{Y}^{\mathrm{UL}}
\mathbf{P}_k^{H}
\approx
\mathbf{H}_k
+
\widetilde{\mathbf{Z}}_k^{\mathrm{UL}},
\end{equation}
where
$\widetilde{\mathbf{Z}}_k^{\mathrm{UL}}
=
\mathbf{F}_{\mathrm{eq}}
\mathbf{Z}^{\mathrm{UL}}
\mathbf{P}_k^{H}$
and
$\mathbf{G}_k\in
\mathbb{C}^{N_{\mathrm{BS}}\times N_{\mathrm{UE}}}$
is the structured noisy UL observation.
Hence, $\mathbf{R}_k$ and $\mathbf{G}_k$ provide unified inputs for learning-based \ac{CSI} refinement, allowing the same compact architecture to be applied to both DL and UL.

\section{Problem Formulation}

\subsection{Federated Channel Estimation Objective}
Based on the DL/UL observation construction in Section~II, each client obtains a structured noisy observation for channel estimation. Since these observations are locally generated and naturally distributed across users, we formulate \ac{CSI} estimation as a federated supervised learning problem \cite{mcmahan2017communication}.

For notational simplicity, let $\mathbf{X}_k$ denote the noisy structured observation and let $\mathbf{Y}_k$ denote the corresponding clean channel target, where
\begin{equation}
(\mathbf{X}_k,\mathbf{Y}_k)=
\begin{cases}
(\mathbf{R}_k,\mathbf{H}_k^H), & \text{for DL},\\
(\mathbf{G}_k,\mathbf{H}_k), & \text{for UL}.
\end{cases}
\end{equation}
To fit the neural estimator, each complex-valued matrix is converted into a real-valued tensor by stacking its real and imaginary parts along the channel dimension. This unified representation enables the same compact estimator design to be applied consistently to both downlink and uplink refinement.

Let $N_k$ denote the number of local training samples at \ac{UE} $k$. The local dataset is
\begin{equation}
\mathcal{D}_k=\{(\mathbf{X}_k^{(i)},\mathbf{Y}_k^{(i)})\}_{i=1}^{N_k}.
\end{equation}
Given a neural estimator $f_{\boldsymbol{\eta}}(\cdot)$, the local training objective is defined as
\begin{equation}\label{eq:local_loss_funtion_MSE}
\mathcal{L}_k(\boldsymbol{\eta})=
\frac{1}{N_k}\sum_{i=1}^{N_k}
\left\|
\mathbf{Y}_k^{(i)}-f_{\boldsymbol{\eta}}(\mathbf{X}_k^{(i)})
\right\|_F^2.
\end{equation}
The corresponding global objective is
\begin{equation}\label{eq:global_obj}
\min_{\boldsymbol{\eta}} \sum_{k=1}^{K} p_k \mathcal{L}_k(\boldsymbol{\eta}),
\end{equation}
where $p_k=\frac{N_k}{\sum_{j=1}^{K} N_j}$ is the aggregation weight of \ac{UE} $k$, typically chosen according to the local data size.

\subsection{Communication-Constrained Federated Optimization}
The federated setting avoids uploading raw \ac{CSI} samples to the \ac{BS}, but each communication round still requires global model broadcast and local model upload. This overhead becomes critical when the estimator is exchanged repeatedly over many rounds, making communication efficiency closely tied to the number of transmitted parameters.

In round $t$, the BS broadcasts the current global parameters to the participating client set $S_t$, and each selected client uploads its updated local parameters to the \ac{BS}. Let $P_{\mathrm{model}}$ denote the number of transmitted parameters, and let $b_{DL}$ and $b_{UL}$ denote the downlink and uplink bitwidth per parameter, respectively. We use a first-order communication accounting model in which the dominant per-round cost is
\begin{equation}\label{eq:bits_consumption_per_rounds}
B_{\mathrm{round}}(t)=|S_t|\,P_{\mathrm{model}}\,(b_{DL}+b_{UL}).
\end{equation}
This accounting captures the leading dependence on parameter count under repeated model exchange while abstracting away protocol overhead and implementation-specific details. Under standard FP32 transmission, $b_{DL}$ and $b_{UL}$ are fixed, so \eqref{eq:bits_consumption_per_rounds} shows that the main estimator design variable that can be reduced is the parameter count. This motivates a compact channel estimator for communication-constrained \ac{FL}.

Given a total communication budget $B_{\max}$, the learning problem is formulated as
\begin{equation}\label{eq:comm_constrained}
\min_{\boldsymbol{\eta}} \sum_{k=1}^{K} p_k \mathcal{L}_k(\boldsymbol{\eta})
\quad \text{s.t.}\quad \sum_{t=1}^{T} B_{\mathrm{round}}(t)\le B_{\max}.
\end{equation}
The main objective of this work is not to reduce the number of communication rounds through a new federated optimization rule, but to study how estimator design affects the \ac{NMSE}--communication tradeoff under an explicit communication budget. Accordingly, in addition to the final \ac{NMSE}, we also evaluate the cumulative communication required to reach a target estimation accuracy. In the following, \ac{BARRNet} is trained under standard \ac{FedAvg}, and the budget-aware aspect is reflected through the communication constraint in \eqref{eq:comm_constrained}, rather than through a new federated optimization or update compression mechanism. Therefore, the goal is not simply to reduce model size, but to use the available model capacity more effectively under a fixed communication budget.

\section{Budget-Aware Recalibrated Refinement Network}

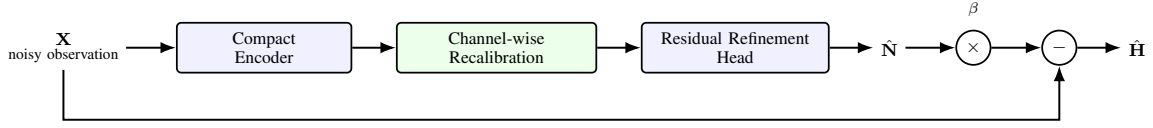
\begin{figure*}[t]
\centering
\resizebox{0.84\textwidth}{!}{%
\begin{tikzpicture}[
    >=Latex,
    font=\footnotesize,
    node distance=5mm and 7mm,
    blk/.style={draw, rounded corners=2pt, minimum height=8mm, align=center, line width=0.8pt},
    cblk/.style={blk, fill=blue!6},
    gblk/.style={blk, fill=green!8},
    op/.style={draw, circle, minimum size=5.5mm, inner sep=0pt, line width=0.8pt}
]

\node[align=center] (x) {$\mathbf{X}$\\[-1mm]\scriptsize noisy observation};
\node[cblk, right=8mm of x, minimum width=28mm] (enc) {Compact\\Encoder};
\node[gblk, right=7mm of enc, minimum width=32mm] (cg) {Channel-wise\\Recalibration};
\node[cblk, right=7mm of cg, minimum width=30mm] (head) {Residual Refinement\\Head};
\node[right=7mm of head] (nhat) {$\hat{\mathbf N}$};
\node[op, right=8mm of nhat] (mul) {$\times$};
\node[above=1mm of mul] {\scriptsize $\beta$};
\node[op, right=8mm of mul] (minus) {$-$};
\node[right=7mm of minus] (hout) {$\hat{\mathbf H}$};

\draw[->, line width=0.9pt] (x) -- (enc);
\draw[->, line width=0.9pt] (enc) -- (cg);
\draw[->, line width=0.9pt] (cg) -- (head);
\draw[->, line width=0.9pt] (head) -- (nhat);
\draw[->, line width=0.9pt] (nhat) -- (mul);
\draw[->, line width=0.9pt] (mul) -- (minus);
\draw[->, line width=0.9pt] (minus) -- (hout);

\draw[->, line width=0.9pt] (x.south) -- ($(x.south)+(0,-8mm)$) -| (minus.south);

\end{tikzpicture}%
}
\caption{Overview of \ac{BARRNet}. A compact encoder $\mathcal{E}_{\boldsymbol{\phi}}$ extracts latent features, a channel-wise recalibration module $\mathcal{G}_{\boldsymbol{\psi}}$ reweights them, and a residual refinement head $\mathcal{D}_{\boldsymbol{\omega}}$ predicts the disturbance estimate $\hat{\mathbf N}$ for channel reconstruction $\hat{\mathbf H}=\mathbf X-\beta\hat{\mathbf N}$.}
\label{fig:barrnet_architecture}
\end{figure*}

\subsection{Design Motivation}
In the considered federated setting, every additional parameter contributes directly to repeated model transmission. Under fixed-precision exchange, model size is therefore not merely a complexity issue, but an estimator level communication budget variable.

For the structured dual-side observations considered here, a compact residual backbone already provides a strong baseline. Our hypothesis is that, under repeated model exchange, a small number of extra parameters may be used more effectively for selective feature recalibration than for uniform backbone scaling. This is later tested against an architecture-matched backbone-only control in which the recalibration module is removed while the refinement backbone is otherwise unchanged.

\subsection{Selective Feature Recalibration under a Parameter Budget}
As shown in Fig.~\ref{fig:barrnet_architecture}, the proposed model follows a compact encoder--recalibration--refinement pipeline. Given a structured noisy observation $\mathbf{X}$, where $\mathbf{X}=\mathbf{R}_k$ for downlink estimation and $\mathbf{X}=\mathbf{G}_k$ for uplink estimation, the encoder first extracts latent features
\begin{equation}
\mathbf{F}=\mathcal{E}_{\boldsymbol{\phi}}(\mathbf{X}).
\end{equation}
The channel-wise recalibration module follows the squeeze-and-excitation principle \cite{hu2018squeeze}. Given encoder features $\mathbf{F}\in\mathbb{R}^{C\times H\times W}$, global average pooling is applied to obtain a channel descriptor $\mathbf{s}\in\mathbb{R}^{C}$. A lightweight bottleneck MLP then generates channel-wise gates
\begin{equation}
\mathbf{g}=\sigma\!\left(\mathbf{W}_2\,\delta(\mathbf{W}_1\mathbf{s})\right),
\qquad
\mathbf{F}'=\mathcal{G}_{\boldsymbol{\psi}}(\mathbf{F})=\mathbf{g}\odot\mathbf{F},
\end{equation}
where $\delta(\cdot)$ and $\sigma(\cdot)$ denote ReLU and sigmoid, respectively, and $\odot$ denotes channel-wise multiplication.
The recalibrated representation is passed to a compact residual refinement head
\begin{equation}
\hat{\mathbf{N}}=\mathcal{D}_{\boldsymbol{\omega}}(\mathbf{F}'),
\end{equation}
which predicts the structured disturbance component. The final channel estimate is reconstructed by
\begin{equation}
\hat{\mathbf{H}}=\mathbf{X}-\beta \hat{\mathbf{N}},
\label{eq:lightgate_residual}
\end{equation}
where $\beta$ is a learnable residual scaling factor. The full parameter set is
\begin{equation}
\boldsymbol{\eta}=\{\boldsymbol{\phi},\boldsymbol{\psi},\boldsymbol{\omega},\beta\}.
\end{equation}

% \begin{algorithm}[t]
% \caption{Federated Training of the Proposed Budget-Aware Recalibrated Refinement Network (BARRNet)}
% \label{alg:barrnet_\ac{FedAvg}}
% \begin{algorithmic}[1]
% \STATE Initialize global parameters \\
% $\boldsymbol{\theta}^{(0)}=\{\boldsymbol{\phi}^{(0)},\boldsymbol{\psi}^{(0)},\boldsymbol{\omega}^{(0)},\alpha^{(0)}\}$ at the BS.
% \FOR{$t=0,1,\dots,T-1$}
%     \STATE The BS selects the participating client set $S_t$ and broadcasts $\boldsymbol{\theta}^{(t)}$.
%     \FOR{each client $k\in S_t$ \textbf{in parallel}}
%         \STATE Perform $E$ local training epochs on $\mathcal{D}_k$ and obtain $\boldsymbol{\theta}_k^{(t+1)}$.
%         \STATE Upload $\boldsymbol{\theta}_k^{(t+1)}$ to the BS.
%     \ENDFOR
%     \STATE Aggregate
%     \[
%     \boldsymbol{\theta}^{(t+1)}
%     =
%     \sum_{k\in S_t} p_k \boldsymbol{\theta}_k^{(t+1)},
%     \]
%     where $p_k=\frac{|\mathcal{D}_k|}{\sum_{j\in S_t}|\mathcal{D}_j|}$.
% \ENDFOR
% \end{algorithmic}
% \end{algorithm}

\section{Simulation Results}

\begin{table*}[t]
\centering
\caption{Communication-efficiency summary at SNR = 5 dB ($K=4$). Lower is better.}
\label{tab:result_summary}
\renewcommand{\arraystretch}{1.05}
\setlength{\tabcolsep}{5pt}
\footnotesize
\begin{tabular}{lccccc}
\hline
Method & Params & UL & Bits/round (Gbits) & Best DL/UL NMSE (dB) & Bits to $-12 / -13$ dB (Gbits) \\
\hline
DnCNN         & 558,211   & FP32  & 0.143 & -13.38 / -12.77 & 4.72 / 11.29 \\
Heavy CNN     & 1,077,275 & FP32  & 0.276 & -13.31 / -12.61 & 2.48 / 7.45 \\
Backbone-only & 408,003   & FP32  & 0.104 & -13.89 / -12.88 & 0.84 / 2.72 \\
BARRNet       & 419,027   & FP32  & \textbf{0.107} & \textbf{-14.43 / -13.31} & \textbf{0.64 / 1.93} \\
Heavy CNN     & 1,077,275 & 8-bit & 0.172 & -13.28 / -12.59 & 1.72 / 4.65 \\
BARRNet       & 419,027   & 8-bit & \textbf{0.067} & \textbf{-14.39 / -13.30} & \textbf{0.40 / 1.27} \\
\hline
\end{tabular}
\end{table*}

\subsection{Experimental Setup}
We evaluate \ac{BARRNet} in a hybrid mmWave massive MIMO system with $K=4$, $N_{\mathrm{BS}}=64$, $N_{\mathrm{RF}}=16$, and $N_{\mathrm{UE}}=4$. The channel model uses $L=3$ dominant paths and \acp{ULA} with half-wavelength spacing. For uplink preamble construction, the \ac{BS} employs $D=4$ combining stages and pilot length $T=16$, such that $D N_{\mathrm{RF}}=64=N_{\mathrm{BS}}$. 

The training pairs are $(\mathbf{R}_k,\mathbf{H}_k^H)$ for DL and $(\mathbf{G}_k,\mathbf{H}_k)$ for UL. \ac{LMMSE} is included as a classical reference. As external neural baselines, we consider a \ac{DnCNN}-based federated baseline inspired by Han \textit{et al.}~\cite{han2025federated} and instantiated using the canonical \ac{DnCNN} architecture of Zhang \textit{et al.}~\cite{zhang2017beyond}, as well as a larger Heavy \ac{CNN} baseline that follows the same overall refinement pipeline as \ac{BARRNet} but uses a deeper and wider intermediate \ac{CNN} backbone. Backbone-only serves as the primary architecture-matched control by removing the recalibration module while keeping the encoder and refinement head otherwise unchanged, thereby testing whether the gain of \ac{BARRNet} comes from more effective parameter allocation rather than merely increased model size. We further include 8-bit uplink-quantized variants as communication-reduced references. Since the practical \ac{FL} bottleneck is typically the client upload, only uplink model transmission is quantized. Accordingly, ``BARRNet + 8-bit UL'' is treated as a compressed variant of the same estimator rather than as a separate method. For fairness, all neural baselines are retrained under the same data generation process, FedAvg protocol, and communication-accounting setup.

The performance metric is
\begin{equation}
\mathrm{NMSE}
=
\mathbb{E}\!\left[
\frac{\|\hat{\mathbf H}-\mathbf H\|_F^2}{\|\mathbf H\|_F^2}
\right].
\end{equation}

Because DL and UL observations are generated through different preprocessing pipelines, the same nominal \ac{SNR} does not imply the same effective observation quality. To enable a fair DL/UL comparison, we introduce a constant uplink \ac{SNR} offset,
\begin{equation}
\mathrm{SNR}_{UL}=\mathrm{SNR}+\Delta_{\mathrm{SNR}},
\end{equation}
where $\Delta_{\mathrm{SNR}}$ is calibrated so that DL and UL share the same no-learning baseline NMSE. For the considered configuration, $\Delta_{\mathrm{SNR}}=1.53$ dB. We use \ac{SNR} = 5 dB as the representative operating point for communication-to-target comparisons, while additional \ac{SNR} values are reported to assess robustness and dual-side consistency.

\begin{figure}[t] 
\centering 
\includegraphics[width=\columnwidth]{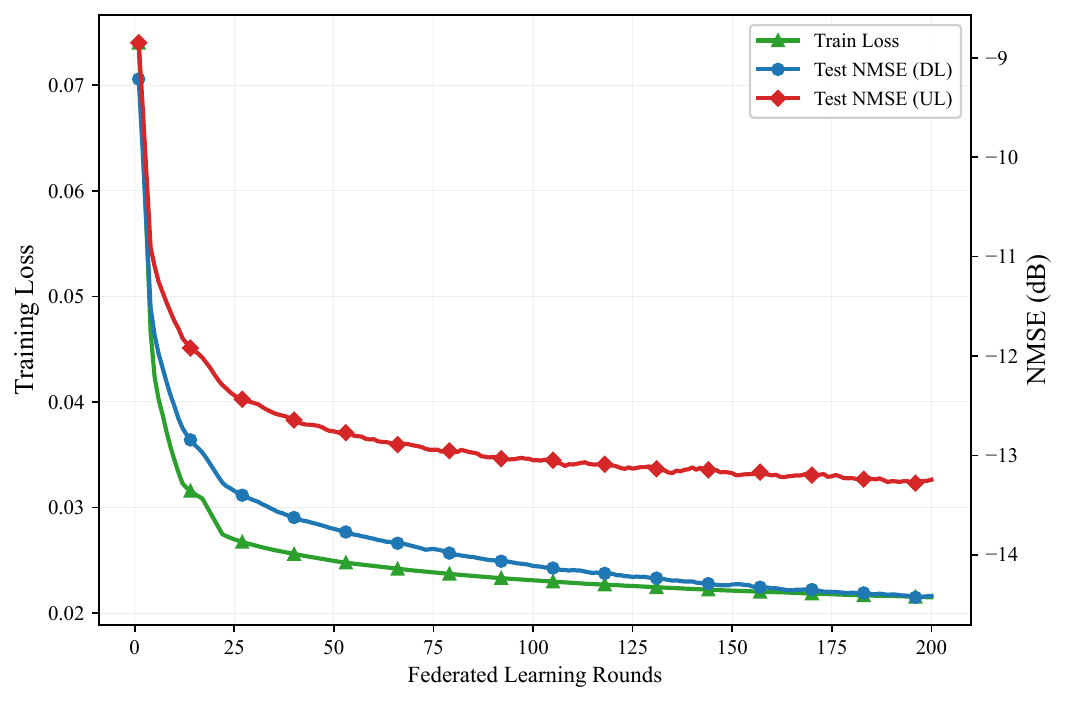} 
\caption{Convergence of the shared \ac{BARRNet} on both DL and UL refinement tasks at SNR = 5 dB. The same model improves both DL and UL NMSE over federated rounds.}
\label{fig:dnqnn_convergence} 
\end{figure}

\subsection{Communication-Efficient Performance Comparison}
Fig.~\ref{fig:dnqnn_convergence} shows that the shared \ac{BARRNet} can be trained stably under \ac{FedAvg} while improving both DL and UL NMSE over federated rounds, supporting the intended dual-side reuse of the same compact refinement architecture. The DL/UL gap is expected because the DL and UL observations are obtained through different preprocessing steps. As a result, the effective noise and reconstruction difficulty are not exactly the same on the two sides, even under the same nominal SNR. 

\begin{figure}[t]
  \centering
  \includegraphics[width=\columnwidth]{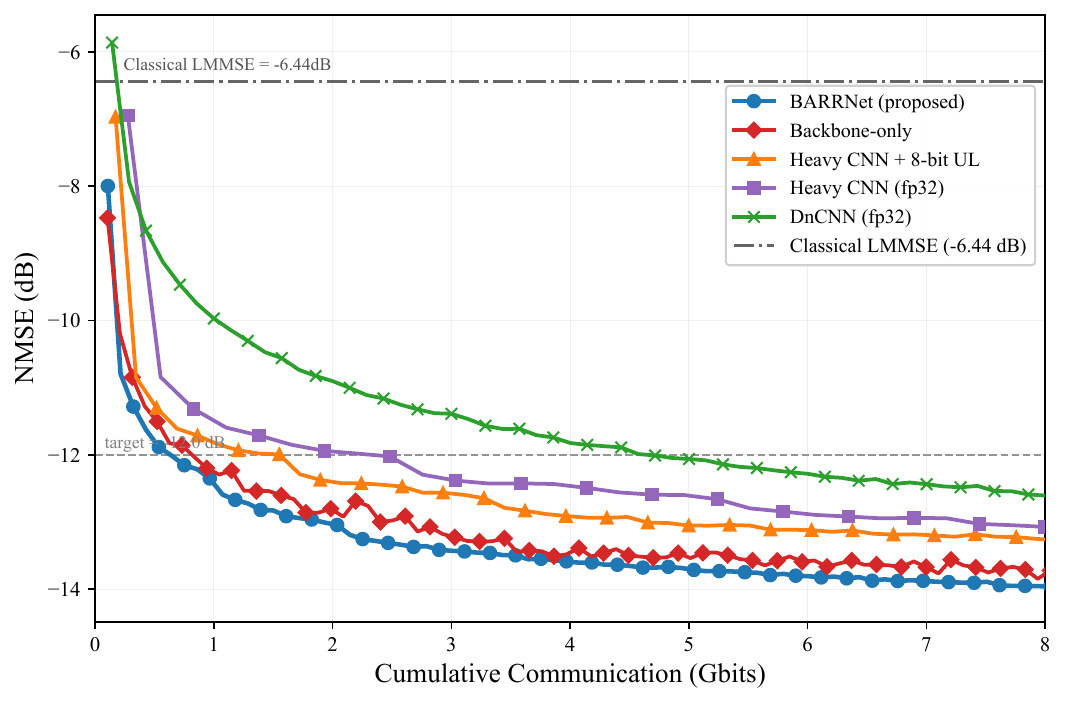}
  \caption{DL NMSE versus cumulative communication cost at SNR = 5 dB.}
  \label{fig:comm_nmse_dl}
\end{figure}

Fig.~\ref{fig:comm_nmse_dl} shows the main communication-efficiency result by comparing DL NMSE against cumulative transmitted bits. Under full-precision model exchange, \ac{BARRNet} achieves the best NMSE--communication tradeoff among the federated baselines. For reference, the horizontal \ac{LMMSE} baseline attains $-6.44$ dB, whereas \ac{BARRNet} reaches $-14.43$ dB.

Comparisons with \ac{DnCNN} and Heavy \ac{CNN} establish the overall competitiveness of \ac{BARRNet}, while the architecture-matched backbone-only control isolates whether the gain comes from more effective parameter allocation rather than merely increased model size. Relative to backbone-only, selective recalibration improves the best DL NMSE from $-13.89$ dB to $-14.43$ dB with only a marginal increase in model size, while reaching the same target accuracy with fewer transmitted bits than both backbone-only and the heavier \ac{CNN} baselines. This supports the design motivation of \ac{BARRNet} that under a limited parameter budget, lightweight recalibration uses model capacity more effectively than uniform backbone scaling.

\begin{table}[t]
\centering
\caption{Budget-allocation ablation at SNR = 5 dB. Here, $(C,N_e,N_h,G)$ denotes the feature width, encoder depth, head depth, and gate hidden width, respectively. Lower is better.}
\label{tab:budget_ablation}
\renewcommand{\arraystretch}{1.03}
\setlength{\tabcolsep}{3.5pt}
\scriptsize
\begin{tabular}{lcccc}
\hline
Variant & $(C,N_e,N_h,G)$ & Params & Best DL/UL (dB) & Bits to $-13$ dB \\
\hline
Balanced       & $(80,2,4,24)$  & 419k & \textbf{-14.43 / -13.31} & \textbf{1.93} \\
Encoder-heavy  & $(76,3,3,12)$  & 429k & -14.00 / -12.97 & 3.40 \\
Head-heavy     & $(76,2,5,12)$  & 429k & -14.06 / -13.09 & 2.52 \\
Recalib-heavy  & $(76,2,4,160)$ & 425k & -14.00 / -13.00 & 3.04 \\
\hline
\end{tabular}
\end{table}

\subsection{Communication-to-Target Cost and Budget Allocation}
\begin{figure}[t]
  \centering
  \includegraphics[width=\columnwidth]{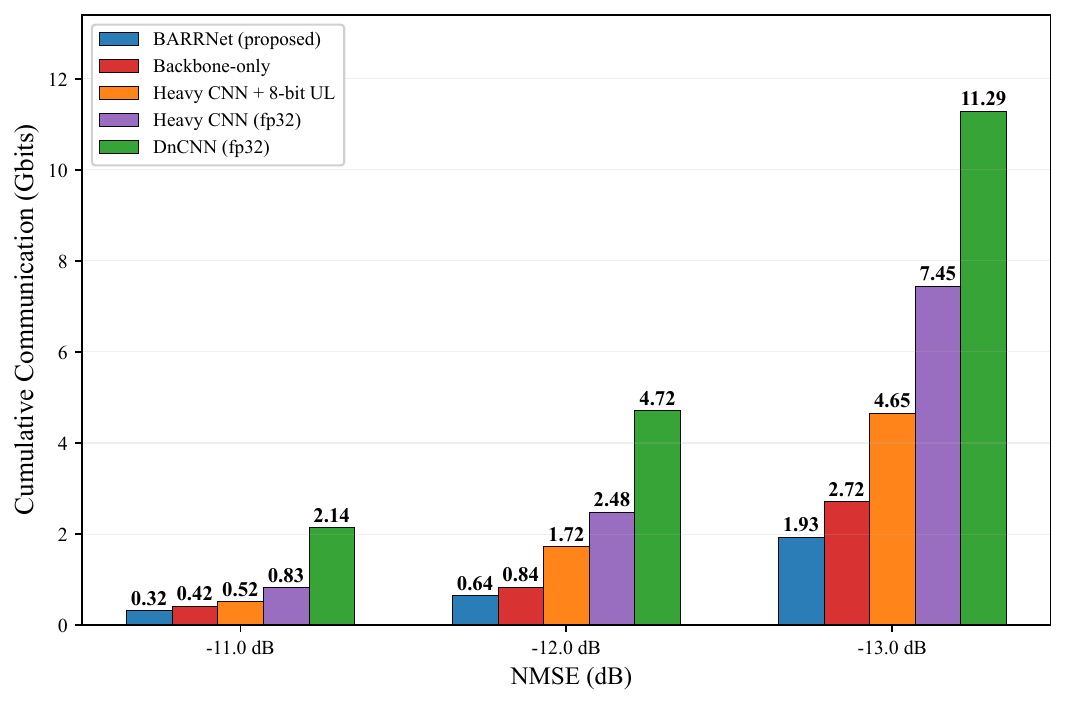}
  \caption{Communication cost required to reach target DL NMSE levels at SNR = 5 dB.}
  \label{fig:bits_target_dl}
\end{figure}

Fig.~\ref{fig:bits_target_dl} reports the cumulative communication required to reach different target DL NMSE levels. Under full precision \ac{FedAvg}, \ac{BARRNet} reaches the same target accuracy with fewer transmitted bits than both the heavier baselines and the architecture-matched backbone-only control. In particular, the communication required to reach -13 dB is reduced from 2.716 Gbits for backbone-only to 1.931 Gbits for \ac{BARRNet}.

Compared with the architecture-matched backbone-only control, \ac{BARRNet} increases the parameter count by only 2.7\% (419,027 vs.\ 408,003), yet improves the best DL NMSE from -13.89 dB to -14.43 dB and reduces the communication required to reach -13 dB from 2.716 to 1.931 Gbits, corresponding to a 28.9\% reduction in communication cost at SNR = 5 dB. Relative to the external DnCNN baseline, \ac{BARRNet} also improves the best DL NMSE from -13.38 dB to -14.43 dB while reducing the communication required to reach -13 dB from 11.289 to 1.931 Gbits. In addition, the 8-bit uplink results show that compact estimator design and classical \ac{FL} communication compression are complementary, since applying uplink quantization preserves nearly the same best DL NMSE while further reducing communication cost.

To test the budget-allocation hypothesis more directly, Table~\ref{tab:budget_ablation} compares four variants with similar parameter budgets but different allocations across the encoder, recalibration module, and refinement head. The balanced allocation used by \ac{BARRNet} achieves the best DL/UL performance ($-14.43/-13.31$ dB) and the lowest communication required to reach the $-13$ dB target (1.93 Gbits). By comparison, the encoder-heavy, head-heavy, and recalibration-heavy variants require 3.40, 2.52, and 3.04 Gbits, respectively, to reach the same target. These results indicate that simply moving more parameters to a single component is less effective than a balanced allocation under the same communication budget.

\subsection{Federated Robustness Across Participation and SNR}
\begin{figure}[t]
  \centering
  \includegraphics[width=\columnwidth]{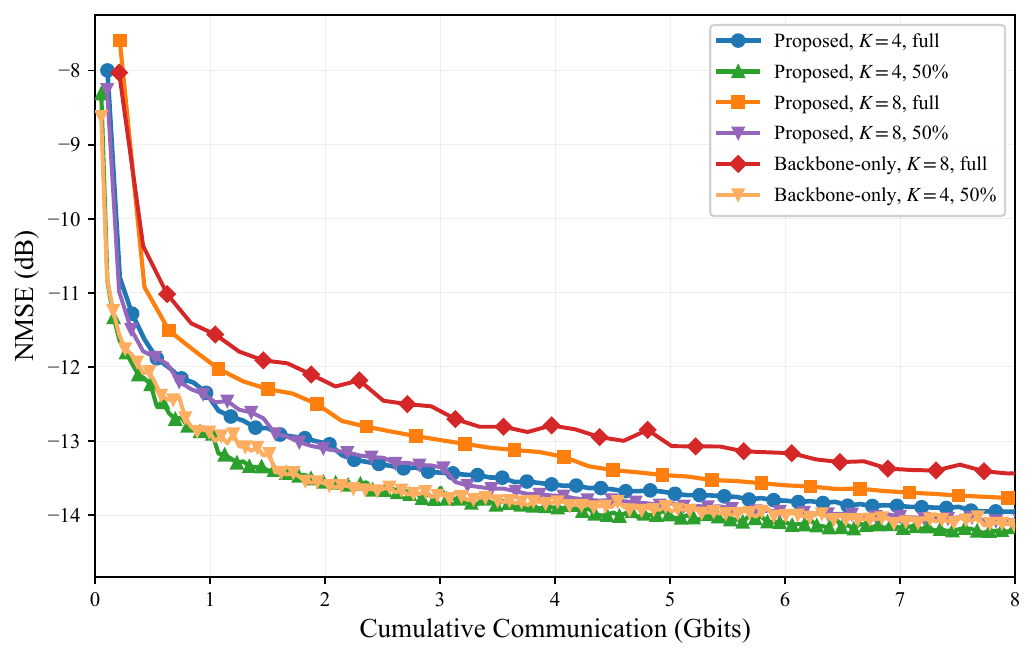}
  \caption{DL NMSE versus cumulative communication cost under different user populations and client participation ratios.}
  \label{fig:larger_participants}
\end{figure}

Fig.~\ref{fig:larger_participants} shows that the communication-efficiency advantage of \ac{BARRNet} is preserved under a larger user population ($K=8$) and under 50\% client participation. Increasing the number of participating users can improve NMSE but also increases the per-round communication cost, whereas partial participation reduces communication and can yield a more favorable operating point. Table~\ref{tab:snr_dualside_nmse} further shows that \ac{BARRNet} remains effective across different SNR regimes on both DL and UL. As the \ac{SNR} increases from 0 to 10 dB, all methods improve, while \ac{BARRNet} consistently achieves the best federated performance and remains close to centralized \ac{BARRNet}, indicating effective training under standard \ac{FedAvg}.

\begin{table}[t]
\centering
\caption{Best converged DL/UL NMSE (dB) across different SNR regimes. Lower is better.}
\label{tab:snr_dualside_nmse}
\renewcommand{\arraystretch}{1.03}
\setlength{\tabcolsep}{3pt}
\scriptsize
\begin{tabular}{lccc}
\hline
Method & 0 dB & 5 dB & 10 dB \\
\hline
Centralized BARRNet  & -10.18/-9.18 & -14.54/-13.65 & -18.74/-17.94 \\
BARRNet              & \textbf{-9.99/-9.03} & \textbf{-14.44/-13.31} & \textbf{-18.41/-17.32} \\
Backbone-only        & -9.57/-8.80 & -13.89/-12.96 & -18.24/-17.11 \\
Heavy CNN + 8-bit UL & -9.11/-8.68 & -13.28/-12.59 & -17.19/-16.35 \\
\hline
\end{tabular}
\end{table}

\section{Conclusion}

This work studied communication-constrained federated \ac{CSI} estimation for hybrid mmWave massive MIMO systems. To enable learning-based refinement under compressed hybrid observations, we designed structured downlink and uplink observation construction procedures and proposed \ac{BARRNet}, which enhances a compact residual backbone through lightweight channel-wise recalibration rather than indiscriminate backbone scaling. The method was developed under standard \ac{FedAvg}, with emphasis on estimator design rather than a new federated optimization algorithm.

Simulation results showed that \ac{BARRNet} achieves a better NMSE--communication tradeoff than both external \ac{CNN} baselines and the architecture-matched backbone-only control. The additional same-budget ablation further indicates that the gain comes not simply from adding parameters, but from allocating a limited parameter budget more effectively. Overall, communication-efficient \ac{FL}-based \ac{CSI} estimation should be studied not only through communication compression, but also through budget-aware estimator design.



%\bibliographystyle{IEEEtran}
%\bibliography{IEEEabrv,../bib/work}

\begin{acronym}
    \acro{FL}{Federated learning}
    \acro{CSI}{channel state information}
    \acro{MIMO}{multiple-input multiple-output}
    \acro{mmWave}{Millimeter-wave}
    \acro{RF}{radio-frequency}
    \acro{BARRNet}{budget-aware recalibration refinement network}
    \acro{NMSE}{normalized mean square error}
    \acro{CNN}{convolutional neural network}
    \acro{FedAvg}{federated averaging}
    \acro{BS}{base station}
    \acro{UE}{user equipment}
    \acro{TDD}{time division duplex}
    \acro{AoA}{angle of arrival}
    \acro{AoD}{angle of departure}
    \acro{LMMSE}{Linear minimum mean square error}
    \acro{DnCNN}{denoising convolutional neural network}
    \acro{SNR}{signal-to-noise ratio}
    \acro{DFT}{discrete Fourier transform}
    \acro{ULA}{uniform linear array}
    \acro{MLP}{Multilayer Perceptron}
    \acro{UL}{uplink}
    \acro{DL}{downlink}
    \acro{VQC}{variational quantum circuit}
\end{acronym}
\bibliographystyle{IEEEtran} 
\bibliography{Ref}

\end{document}